\pdfoutput=1

\documentclass[11pt]{article}

\usepackage[final]{acl}
\usepackage[normalem]{ulem}
\usepackage{cancel}
\usepackage{soul}
\usepackage{amsmath}
\usepackage{multirow}
\usepackage{multicol}
\usepackage{amssymb}
\usepackage{booktabs}

\usepackage{times}
\usepackage{latexsym}

\usepackage[T1]{fontenc}

\usepackage[utf8]{inputenc}

\usepackage{microtype}

\usepackage{inconsolata}

\usepackage{graphicx}
\title{GSID: Generative Semantic Indexing for E-Commerce Product Understanding}

\author{
 \textbf{Haiyang Yang\textsuperscript{1}},
 \textbf{Qinye Xie\textsuperscript{1,}\footnotemark[1]},
 \textbf{Qingheng Zhang\textsuperscript{1,}\footnotemark[1]},
 \textbf{Liyu Chen\textsuperscript{1}},
 \textbf{Huike Zou\textsuperscript{1}},
 \\
 \textbf{Chengbao Lian\textsuperscript{1}},
 \textbf{Shuguang Han\textsuperscript{1,}\footnotemark[2]},
 \textbf{Fei Huang\textsuperscript{1}},
 \textbf{Jufeng Chen\textsuperscript{1}},
 \textbf{Bo Zheng\textsuperscript{1,2}}
\\
 \textsuperscript{1}Xianyu of Alibaba \quad
 \textsuperscript{2}Taobao\&Tmall Group of Alibaba
\\
\textbf{\{yanghaiyang.yhy,qingheng.zqh,shuguang.sh\}@alibaba-inc.com}
}

\begin{document}
\maketitle
\begin{abstract}

\renewcommand{\thefootnote}{\fnsymbol{footnote}}
\footnotetext[1]{These authors contributed equally to this work.}
\footnotetext[2]{Corresponding author}

Structured representation of product information is a major bottleneck for the efficiency of e-commerce platforms, especially in second-hand ecommerce platforms. Currently, most product information are organized based on manually curated product categories and attributes, which often fail to adequately cover long-tail products and do not align well with buyer preference. To address these problems, we propose \textbf{G}enerative \textbf{S}emantic \textbf{I}n\textbf{D}exings (GSID), a data-driven approach to generate product structured representations. GSID consists of two key components: (1) Pre-training on unstructured product metadata to learn in-domain semantic embeddings, and (2) Generating more effective semantic codes tailored for downstream product-centric applications. Extensive experiments are conducted to validate the effectiveness of GSID, and it has been successfully deployed on the real-world e-commerce platform, achieving promising results on product understanding and other downstream tasks.
\end{abstract}

\section{Introduction}
The widespread adoption of digital technologies and the evolution of e-commerce ecosystems have led to rapid growth in the consumer-to-consumer (C2C) second-hand e-commerce domain\cite{wu2025iu4rec}. Unlike the traditional business-to-consumer (B2C) model\cite{chan2023capturing,zhou2018deep}, C2C platforms have a majority of products listed by individual users, often with limited inventory\cite{wu2024metasplit}, while also hosting professional sellers with multiple stock-keeping units. The diversity has resulted in a severe long-tail problem in C2C platforms\cite{wu2025iu4rec}, making it challenging to effectively organize product information.

Traditionally, product understanding has been organized using manually curated product categories and attributes named Category-Property-Values (CPV\cite{su2025taclr}), which not only have limited coverage of long-tail products but also often misalign with buyer preference which is usually manifested through user search queries, as the structure is typically designed from the seller's perspective.
\begin{figure}[t]
  \includegraphics[width=\columnwidth]{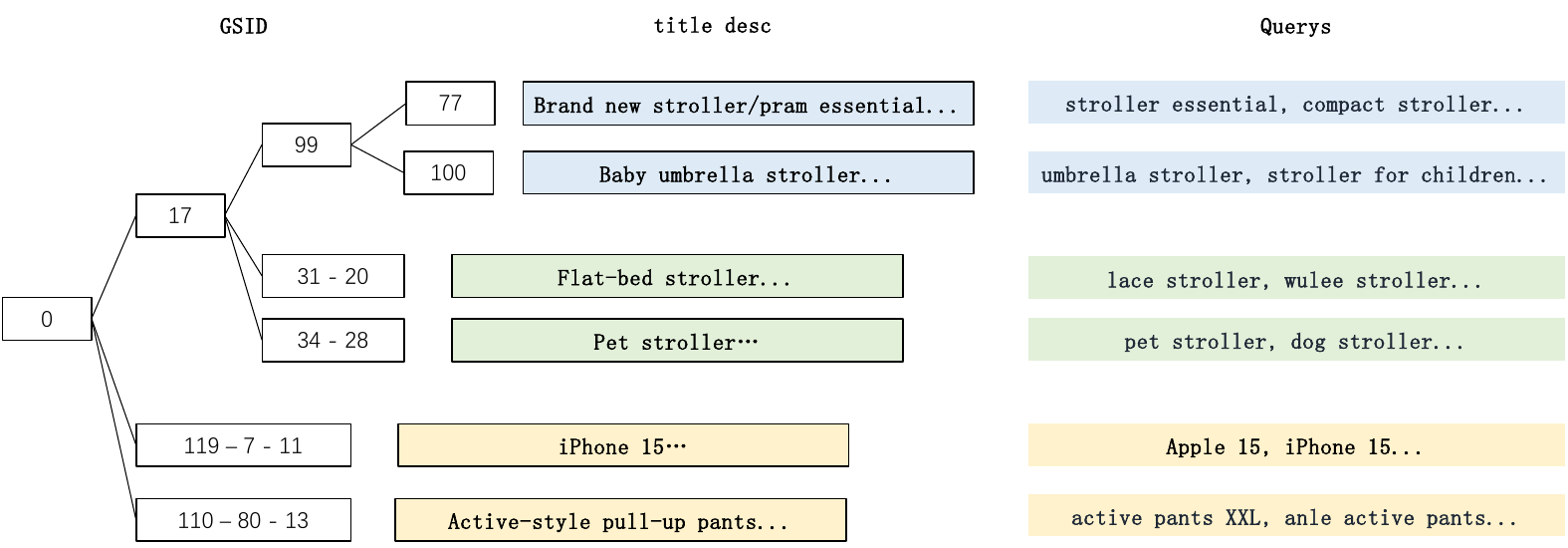}
  \caption{Examples of product descriptions and associated search queries with their GSID.}
  \label{fig:intro}
\end{figure}

Inspired by tokenization learning paradigms\cite{sun2023learning,lee2022autoregressive,tay2022transformer}, we propose a data-driven, self-organizing approach to address the aforementioned issues. Tokenization learning methods typically utilize an encoder model to project unstructured text data into discrete latent spaces, generating discrete, interpretable document representations in an unsupervised framework. Among the various tokenization learning methods, we choose GenRet\cite{sun2023learning}, a novel document tokenization learning method that learns to tokenize documents into short and discrete representations using a discrete auto-encoding framework with progressive training scheme, as our base model, considering its potential for end-to-end performance and efficiency breakthroughs through global optimization, automatic feature learning, and simplified workflows, as well as its hierarchical structure that provides stronger interpretability.

Furthermore, we leverage user queries on e-commerce platforms, which not only reflect users' latest fine-grained needs but also contain user behavior patterns, to guide the structuring of product information, thereby addressing the limitations of traditional CPV-based product organization.
\begin{table}[t]
  \centering
  \begin{tabular}{lcc}
  \toprule
    Cluster&Coverage&Accuracy\\
    \midrule
    CPV&23.02&96.98\\
    GSID&78.81&97.09\\
    \bottomrule
  \end{tabular}
  \caption{\label{tab:intro}
    Comparison between CPV Cluster and GSID Cluster. The accuracy of cluster is manual annotation in evaluation test with a million samples.
  }
\end{table}

However, directly applying GenRet\cite{sun2023learning} to C2C product understanding tasks faces the challenge that GenRet-related algorithms lack successful large-scale industrial applications. Thus we propose \textbf{G}enerative \textbf{S}emantic \textbf{I}n\textbf{D}exing (\textbf{GSID}), a data-driven approach that generates discrete product structured representations tree based on the tokenization learning paradigms as shown in Fig.\ref{fig:intro}, which consists of two key stages: (1) Pre-training on unstructured product metadata to learn in-domain semantic embeddings, and (2) Generating more reliable structured representations tailored for product understanding tasks, leveraging the insights gained from the in-domain pre-training. The proposed improvements not only alleviate the long-tail data coverage issues in products but also generate more reliable structured representations on industrial-scale C2C product data, which can be found in Tab.\ref{tab:intro}. Furthermore, GSID partially solves challenges in downstream applications, such as search and recommendation systems, achieving significant performance improvements compared to previous methods.

Our contributions can be summarized as follows: (1) We introduce a generative semantic indexing framework that establishes a data-driven product understanding paradigm, overcoming the limitations of traditional rule-based systems. This method learns semantic representations directly from raw product data through end-to-end training, eliminating the reliance on manually crafted heuristics. (2) We optimize the GenRet training framework to overcome the challenge of its difficulty in training on large-scale datasets. (3) We demonstrate the effectiveness of GSID in product understanding and downstream tasks, particularly in the second-hand e-commerce domain. Extensive experiments validate the reliability of the proposed method and provide a scalable application example for industrial use.
\section{Methods}
\begin{figure*}[t]
  \includegraphics[width=2\columnwidth]{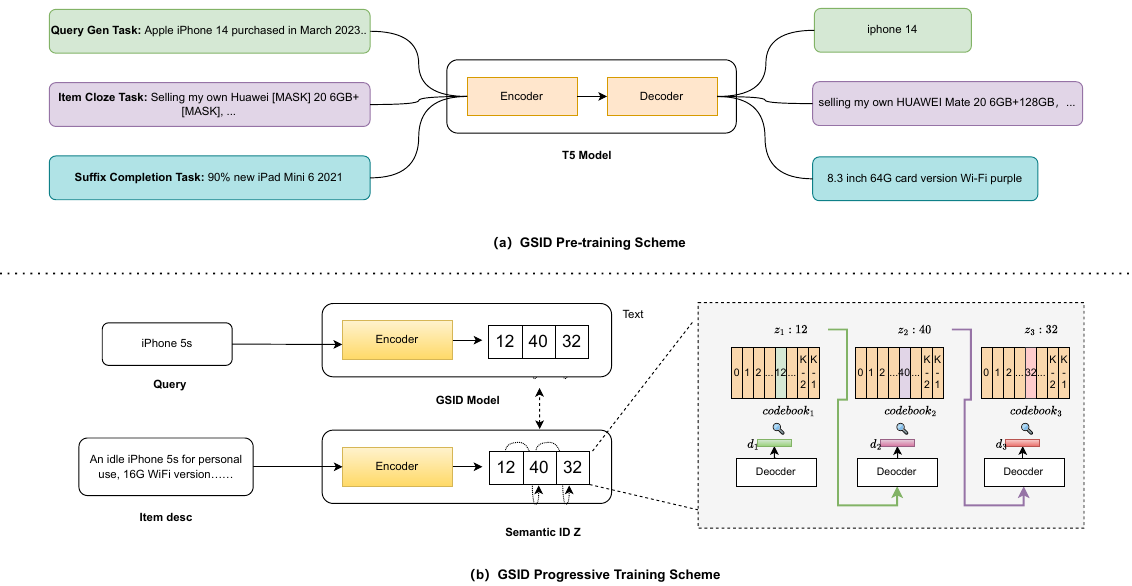}
  \caption{Pipeline of GSID. (a) illustrates the domain-adaptive pre-training scheme. (b) shows the progressive training scheme}
  \label{fig:architecture}
\end{figure*}

In this section, we present Generative Semantic InDexing (GSID), a novel approach to product understanding based on generative semantic indexing. Fig.\ref{fig:architecture} illustrates the overall architecture of our proposed framework.

\subsection{Model Architecture}
Following GenRet, we employ a T5-based encoder-decoder transformer\cite{raffel2020exploring} to implement the generative semantic indexing model. The encoder component of the model is responsible for extracting semantic representations of the input text $d$, which could be product descriptions or search queries. The decoder component progressively decodes and generates semantic IDs, conditioned on the encoder's output and the prefix of previously generated semantic IDs $z_{<t}$. The output of the decoder at step $t$ is a latent representation $d_t$:$$d_t = \text{Decoder}(\text{Encoder}(d), z_{<t}) \in \mathbb{R}^D$$

where $D$ represents the hidden size of the model. Furthermore, at each step $t$, we define an external embedding matrix, namely, a codebook\cite{van2017neural} $E_t \in \mathbb{R}^{K \times D}$, where $K$ is the size of the discrete latent space, which also defines the range of semantic IDs. The discrete latent variable $z_t$ at step t can be obtained by a dot-product lookup against the codebook $E_t$:$$Q(z_t = j| z_{<t}, d) = \text{softmax}_{j}(d_t \cdot E_t^T)$$

where $Q(z_t = j| z_{<t}, d)$ denotes the probability that the semantic IDs for input $d$ at step $t$ is $j \in [K]$. Through this sequential decoding process, the lookup of $d_t$ against the codebook enables vector quantization, thereby generating hierarchical semantic IDs.
\subsection{Training Methodology}
Our training methodology employs a two-stage approach, comprising domain-adaptive pre-training and a progressive training scheme. The pre-training phase aims to inject e-commerce domain knowledge into the open-source T5 model, thereby enhancing its performance in e-commerce scenarios. The progressive training phase focuses on the generation of semantic IDs, resulting in the final GSID.
\subsubsection{Domain-Adaptive Pre-training Scheme}
GSID is built upon Mengzi T5\cite{zhang2021mengzi}, a Chinese pre-trained model. While Mengzi T5 encompasses general domain knowledge, it does not align well with the distribution of e-commerce data. Continue pre-training tasks are thus utilized, taking search queries and product descriptions as input, as illustrated in Figure \ref{fig:architecture}(a). Specifically, we design three training tasks with loss as format $\mathcal{L}_{\text{pretrain}} = \mathcal{L}_{\text{query\_generation}} + \mathcal{L}_{\text{item\_cloze}} + \mathcal{L}_{\text{item\_completion}}$. Each of these tasks utilizes a standard negative log-likelihood loss for sequence generation. Formally, for a given task $\mathcal{T}$, the loss is defined as: $$\mathcal{L}_{\mathcal{T}} = -\sum_{k=1}^{L} \log P(y_k | y_{<k}, \text{Encoder}(\mathbf{x}_{\mathcal{T}}))$$where $\mathbf{x}_{\mathcal{T}}$ represents the input sequence of task $\mathcal{T}$ fed into the T5 encoder, $y = [y_1, y_2, ..., y_L]$ is the target output sequence generated by the decoder, $L$ is the length of the target sequence. The specific inputs and targets for each task are follows:

\textbf{Query Generation Task} The encoder's input $\mathbf{x}_{\text{QueryGen}}$ is a product description $d$, and the target sequence $y$ is the corresponding user query $q$. This task trains the model to understand the semantic intent of a product description.

\textbf{Item Cloze Task} The encoder's input $\mathbf{x}_{\text{ItemCloze}}$ is a masked product description $d_{\text{masked}}$, and the target sequence $y$ is the complete product description $d$. This task enhances the model's ability to comprehend complete product information, improving its contextual understanding of item details.

\textbf{Product Suffix Completion Task} The encoder's input $\mathbf{x}_{\text{SuffixComp}}$ is a product description prefix $d_{\text{prefix}}$, and the target sequence $y$ is the subsequent content (suffix) of the product description $d_{\text{suffix}}$. This task encourages the model to learn the structural and semantic dependencies within product descriptions, crucial for generating coherent product attributes.
\subsubsection{Progressive Training Scheme}
To further enable the model to generate hierarchical semantic IDs, we adopt a Progressive Training Scheme, as illustrated in Fig.\ref{fig:architecture}(b). The training procedure consists of $M$ steps, corresponding to the $M$ semantic IDs ultimately produced. Furthermore, at each step $T \in [M]$, when training semantic IDs $z_T$, the previously produced semantic IDs $z_{<T}$ are kept fixed. At each optimization step, we use the gradient descent with respect to the loss $\mathcal{L}_{\text{gsid}} = \mathcal{L}_{\text{align}} + \mathcal{L}_{\text{com}}$. The alignment loss of query and product $\mathcal{L}_{\text{align}}$ and code commitment loss $\mathcal{L}_{\text{com}}$ are shown as follow:


\textbf{Query-Item Alignment Objective} To extract user-intent semantics from product descriptions, we introduce a query-item alignment objective to jointly learn representations for queries and items, as well as their corresponding semantic IDs. Specifically, suppose $\langle q, d \rangle$ is a relevant pair of a query and a product description, we define the alignment objective as:
$$
   \mathcal{L}_{\text{align}} = -\log\left( \frac{\exp(q_T \cdot d_T)}{\sum_{d^* \sim \mathcal{B}} \exp(q_T \cdot d^*_T)} \right) 
   $$$$+ \sum_{t=1}^{T} D_{KL}\left( P(z_t | z_{<t}, q) || P(z_t | z_{<t}, d) \right)
$$
where the first term is the contrastive learning loss for the $\langle q, d \rangle$ pair, and $d^*$ is an in-batch negative sample from the same training mini-batch $\mathcal{B}$. $q_T$ and $d_T$ denote the representations of the query and item, respectively, at step $T$. The second term is the KL divergence\cite{perez2008kullback} (by avoiding the use of hard-label cross-entropy loss) between the probability distributions of the query semantic IDs and item semantic IDs, which helps stabilize the training process. 

\textbf{Code Commitment Objective} To prevent the model from forgetting the previously generated semantic IDs $z_{<t}$ when predicting the current step's semantic IDs, we introduce the code commitment loss as follows:
$$\mathcal{L}_{\text{com}} = -\sum_{t=1}^{T} \log Q(z_t | z_{<t}, \text{Encoder}(d))$$

For the codebook update, to improve the stability of codebook training, we do not employ the original VQ-VAE\cite{van2017neural} codebook training algorithm. Instead, we update the codebook using the Exponential Moving Averages (EMA) method\cite{klinker2011exponential}. Given the massive scale of our training data, our codebook employs random initialization instead of the classical constrained K-Means clustering approach\cite{hartigan1979algorithm}. 

\subsubsection{Data Sample Techniques}
Effective training within a Progressive Training Scheme heavily relies on robust data sampling. To further enhance model performance, we adopt two key strategies: 


\textbf{Multi-Query Sampling} In the e-commerce domain, frequently searched queries are often generic terms. These queries are characterized by an enormous number of individual requests and broad semantic meanings (e.g., "mobile phone", "laptop computer"). In such scenarios, it is highly prone to encountering samples of the same concept within mini-batch, which is detrimental to model training. To address this, we sample multiple positive query examples for each item to mitigate the potential impact of false negative samples. 

\textbf{Progressive Negative Sampling} The construction of negative samples is crucial in contrastive learning. Negative samples within a mini-batch are often simple, which the model can easily distinguish, leading to insufficient learning. To increase the difficulty of model training, we adopt a difficulty-increasing adaptive negative sampling method. Specifically, we require that training samples at each step $T$ share the same semantic IDs prefix. Since semantic IDs with the same prefix typically possess similar semantics, negative samples within this range are correspondingly harder for the model, which is more conducive to the model learning fine-grained semantic distinctions.
\section{Experimental Setup}

\subsection{Dataset}
We collect search logs of 20 days from Xianyu\footnote{Xianyu is a C2C e-commerce platform in Alibaba} search platform to build the training dataset with 200 millions samples. To evaluate the effectiveness and generalization of GSID, we conducted a series of experiments on two large-scale real-world datasets collected: Xianyu-Retrieval and Xianyu-Product-Understanding to validate the effectiveness of GSID.

\textbf{Xianyu-Retrieval} Based on Xianyu online click data, we collected query-product pairs (q2i) and product-product pairs (i2i), and ensuring that not share any common pairs with training set. The q2i contain of 30 thousands query triggers and a million product candidates, while the i2i insist of 30 thousands product triggers and 150 thousands product candidates.

\textbf{Xianyu-Product-Understanding} From Xianyu item pool, we sampled a million products with their operationally-annotated CPV information. 

\begin{table*}[t]
  \centering
  \begin{tabular}{l|cccc|cccc}
    \toprule
    \multirow{2}{*}{\textbf{Methods}}     & \multicolumn{4}{c|}{\textbf{Q2I}} &\multicolumn{4}{c}{\textbf{I2I}} \\
    & \textbf{R@1} & \textbf{R@10} &\textbf{R@100}&\textbf{MRR@100}& \textbf{R@1} & \textbf{R@10} &\textbf{R@100}&\textbf{MRR@100} \\ 
    \midrule
    BM25       & 22.41           & 44.52&64.11&29.89  & 5.92           & 12.89&19.19&8.24                      \\
    BGE     & 25.96         & 53.07&76.73&35.20  & 8.37           & 19.91&32.48&12.23\\
    XY-BERT       & 26.80           & 55.92&77.69&36.73  & 7.05           & 20.85&33.57&11.82\\
    GSID(ours) & \textbf{31.26}           & \textbf{58.63}&\textbf{77.95}&\textbf{40.60} & \textbf{8.43}           &\textbf{21.88}&\textbf{35.01}&\textbf{13.17}\\
    \bottomrule
  \end{tabular}
  \caption{\label{tab:1}
    Results of query-to-product (q2i) retrieval and product-to-product (i2i) retrieval task.
  }
\end{table*}

\subsection{Baselines}
\label{baselines}
To evaluate the superiority of GSID, we selected several strong baselines for comparison, including:
\begin{itemize} 
    \item \textbf{BM25\cite{robertson1994some}:} a probabilistic model that estimates the probability of a document being relevant to a query, based on the occurrence of query terms in the document.
    \item \textbf{BGE\cite{xiao2024c}} a dense retrieval model that leverages graph neural networks to learn effective representations for documents and queries.
    \item  \textbf{XY-BERT} a bert encoder that trained by masked language modeling, and then finetuned with contrastive learning on Xianyu dataset, which widely used in Xianyu search business.
\end{itemize}

We conducted two types of experiments using these baselines. For the first type, we compared the retrieval performance of using the final representation output by the GSID decoder against these retrieval baselines. And for the other, we constructed semantic encodings using hierarchical clustering on top of these baseline models, and directly compared them against the GSID approach. As BM25 is a pure text matching method, we excluded it from the second type of experiments.

\subsection{Experimental Settings}
To evaluate the semantic similarity modeling capability of GSID, we utilize the q2i and i2i retrieval task on the Xianyu-Retrieval dataset with $Recall@k$. And we evaluate the informative and interpretable nature of GSID through alignment with GSID-Query and GSID-Category relationships which we use AMI and code\_accuracy as the metric on the Xianyu-Understanding dataset. We will provide detailed descriptions of our experimental setup in the Appendix.

\section{Results}
\subsection{Retrieval Results}
In Tab.\ref{tab:1}, we present the results on the Xianyu-retrieval dataset for the two tasks of query-product retrieval (q2i) and product-product retrieval (i2i), to demonstrate the semantic similarity representation capacity of the embeddings. And GSID outperforms all the baselines on both of these tasks. In particular, for recall, GSID can achieve significant improvements. For example, GSID get R@1 of 31.26\% on q2i task, which is \textbf{+4.46\%} better than the best baseline. This result highlights that the representations conducted by GSID have stronger semantic similarity expression capabilities than the baselines. 

\subsection{Interpretability}
Tab.\ref{tab:2} shows the performance comparison between our GSID and other semantic IDs obtained through hierarchical clustering of baseline methods. Cate1\_AMI and cate4\_AMI represent the AMI between the semantic IDs and the Idle Fish category labels at the first and fourth levels, respectively. GSID  achieves a \textbf{+3.09\%} performance improvement on cate1\_AMI and \textbf{+0.47\%} improvement on cate4\_AMI over the best baseline. As for the query-item code consistency, which measures the semantic similarity of GSID, we get 89.48\% on l1\_code\_acc and 70.19\% on l2\_code\_acc, which significantly outperforms other methods. Compared with Tab.\ref{tab:1}, results of Tab.\ref{tab:2} indicate that the final output GSID exhibits superior semantic encoding performance compared to the representations.

\begin{table}
  \centering
  \begin{tabular}{l|ccc}
    \toprule
    \textbf{Methods}&BGE&XY-BERT&GSID(ours)\\ 
    \midrule
    cate1\_AMI & 54.80&56.11&\textbf{59.20}\\
    cate4\_AMI & 41.20&45.33&\textbf{45.80}\\
    l1\_code\_acc & 34.37 &47.50 &\textbf{89.48} \\
    l2\_code\_acc &10.24 &20.31 &\textbf{70.19}\\
    \bottomrule
  \end{tabular}
  \caption{\label{tab:2}
    Results of AMI and code consistency.
  }
\end{table}

\subsection{Visualization}
\begin{figure}[t]
  \includegraphics[width=\columnwidth]{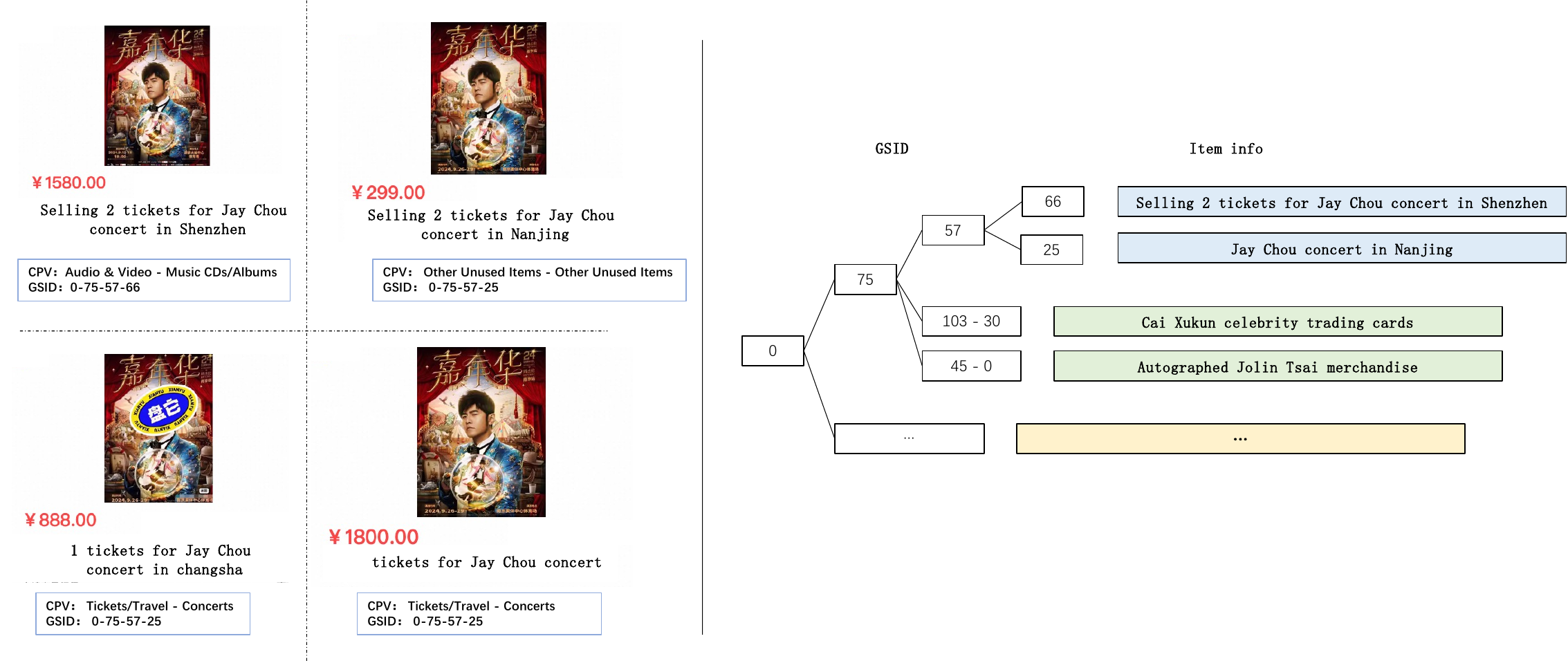}
  \caption{Comparison of CPV and GSID.}
  \label{fig:visual}
\end{figure}

To provide a more intuitive visual comparison of the structural properties between GSID and CPV, we utilize concert tickets for the artist Jay Chou as illustrative examples in the Fig.\ref{fig:visual}. The result shows that, in comparison to CPV, GSID exhibits a finer granularity and an architectural design more aligned with user need. Further, the result also shows that CPV may have given different structured information to the same goods due to operational errors, but GSID can give the same semantic ID to goods with the same semantics because it is data-driven.
\section{Online A/B Test}
The proposed GSID has been successfully intergrated into key functionalities of the e-commerce platform Xianyu, including generative retrieval and user interest modeling. Furthermore, we achieved quite satisfactory online A/B testing gains.

GSID for generative retrieval consists of several components. For the query side, we restrict the beam search\cite{brown1993mathematics,sutskever2014sequence} to generate multiple candidates, while also collecting statistics on the user clicks on these candidates indexing. At the same time, we select high-conversion products based on the GSID. After that, we can perform query-product matching. In generative retrieval, GSID leads to a 1\% increase in GMV and a 0.3\% increase in average transactions per user, while satisfying the relevance constraints. This demonstrates the advantages of GSID's consistent encoding of queries and items, as well as the benefits of end-to-end ID-based modeling.

GSID for user interest modeling means that we regard GSID as a kind of feature in CTR model. We used CTR, Clicks, and Bills as metrics to evaluate the online effectiveness. The IU-Boosted model\cite{wu2025iu4rec} demonstrated significant improvements in both interest unit and normal product recommendations. Specifically, for interest unit recommendations, there was an 11.76\% increase in CTR, a 10.22\% rise in Clicks, and a 5.61\% boost in Bills. For normal product recommendations, CTR improved by 3.72\% and Clicks by 5.20\%. Overall, the model achieved a 2.90\% increase in CTR, a 2.62\% rise in Clicks, and a 1.02\% increase in Bills, highlighting a notable performance enhancement.
\section{Related Work}
\textbf{Product structured representation} The structured understanding of products is a fundamental problem in the e-commerce domain. Early approaches, such as \cite{gao2020deep}, utilized hierarchical category systems to organize products, with deep hierarchical classification often employed for product category prediction. Works like \cite{li2023attgen, zhang2022oa, xu2023towards} introduced attribute tree systems, which are typically constructed via data mining techniques to model richer product information. However, the management and maintenance of these attribute trees still heavily rely on manual expert curation, leading to high operational costs in practical e-commerce scenarios. Further advancing product representation, methods like \cite{luo2020alicoco, li2020alimekg, luo2021alicoco2, dong2019building} leveraged knowledge graphs. These approaches often involve extensive data mining and graph construction efforts to build comprehensive product profiles, which, while significantly enriching the available information, also introduce considerable system complexity.

\textbf{Discrete representation learning} Discrete representation learning is a crucial area in machine learning, especially for efficient data indexing and retrieval. Early works VQ-VAE\cite{van2017neural}, which learn quantized representations via vector quantization, primarily for images. For discrete identifier generation in information retrieval, methods like DSI\cite{tay2022transformer} employ a two-stage process, typically involving clustering over continuous item representations. This two-stage approach, however, suffers from a strong dependency on initial representation quality, and  potential information loss due to the disjointed nature of learning and ID generation. To mitigate these issues, RQ-VAE\cite{lee2022autoregressive} integrates hierarchical quantization end-to-end. GenRet\cite{sun2023learning} employs a T5-based seq2seq architecture to generate discrete codes. Its key innovations include step-specific codebooks for rich information preservation and per-step hard negative sampling.
\section{Conclusion}
In this paper, we proposed a generative semantic indexing framework that establishes a data-driven product understanding paradigm, overcoming the limitations of traditional rule-based systems. This method learns semantic representations directly from raw product data through end-to-end training, eliminating the reliance on manually crafted heuristics. Extensive experiments validate the reliability of the proposed method and provide a scalable application example for industrial use. 
\section{Limitation}
While GSID demonstrates effectiveness in processing textual product profiles, it does not currently leverage multimodal information, such as
product images or videos. Multimodal data could provide valuable complementary context for attributes that are challenging to infer from text alone. Incorporating multimodal capabilities may further enhance the model’s ability.

Another limitation is that once the GSID structure is fixed, it is difficult to make adjustments, which is not conducive to the update of new products. Therefore, how to periodically update the GSID system is an issue that still needs to be explored.

\bibliography{custom}

\appendix

\section{Evaluation Metrics}
\label{sec:appendix}

$Recall@k$ is a metric that measures the proportion of expected product retrieved by the search system. For a given k, $Recall@k$ is defined as: $$Recall@k=\frac{1}{|Q|}\sum_{q=1}^{|Q|}\frac{ret_{q,k}}{rel_q}$$
where $|Q|$ is the number of queries in the set, $ret_{q,k}$ is the number of expected products retrieved for the q-th query within the top k results, and $rel_q$ is the total number of expected documents for q-th query. 

Query-item consistency rate is used to evaluate the alignment between query and item semantic IDs. Similar with Recall@k, we propose l1\_code\_acc and l2\_code\_acc to show the proportion of queries and their retrieved products that share identical primary and secondary IDs. We randomly sampled 150,000 query-item pairs from the q-i dataset to examine whether their GSID distributions are identical. Specifically,
l1\_code\_acc indicates the proportion where Level-1 semantic IDs match and l2\_code\_acc indicates the proportion where both Level-1 and Level-2 semantic IDs are fully consistent;

Mean Reciprocal Rank ($MRR@k$) measures the rank of the correct product category within the top k ranked predictions. For a given k, MRR@k is defined as: $$MRR@k = \frac{1}{|Q|}\sum_{i=1}^{|Q|}\frac{1}{rank_i}$$
where $|Q|$ is the number of queries in the set, $rank_i$ is the rank for i-th item.

Adjusted Mutual Information (AMI) measures the agreement between the predicted product categories and the ground-truth labels, while accounting for the expected level of agreement due to chance. AMI can be defined as: $$AMI(U,V)=\frac{MI(U,V)-E\{MI(U,V)\}}{F(H(U),H(V))-E\{MI(U,V)\}}$$
where $MI(U,V)$ is the Mutual Information of distribution $U$ and $V$, $H(\cdot)$ is the information entropy, $E$ is the expectation and $F(\cdot)$ means a function which we use $max(\cdot)$ in this paper. Furthermore, we use the cate1\_AMI and cate4\_AMI to measure the matching extent between category in CPV with the semantic IDs.

As mentioned before, there are two different experiments to demonstrate the effectiveness of GSID. For the first type of experiment, we selected $recall@\{1, 10, 100\}$ and $MRR@100$ to evaluate the retrieval performance of the representations, following previous works\cite{sun2023learning}. As for the second type, we choose the category Adjusted Multual Information (AMI)\cite{vinh2009information} to show the interpretability of GSID compared to the baselines.
We use $Recall@\{1,10,100\}$ and Mean Reciprocal Rank
(MRR)@100 as evaluation metrics to evaluate the retrieval performance. And we use query-item and item-item ID consistency as the encoding accuracy. We also prove the effectiveness of encoding interpretability with category Adjusted Mutual Information (AMI) and SPU purity within a cluster.

\section{Implementation Details}

\textbf{Dataset Description}

Query-Item Dataset (200 million samples): Constructed based on user interactions with items under different queries, including transaction, interaction, and click behaviors along with their frequencies. Each item corresponds to multiple queries.
Item-Item Dataset (78 billion samples): Obtained by calculating swing I2I from 30-day user click sequences. 1.5 billion samples are extracted from this dataset to form the evaluation set.

\textbf{Domain-Adaptive Pretraining Phase}

We first train the mengzi\_t5 base model using 32 A800 GPUs on the 200-million-sample query-item dataset for 2 epochs. And then we utilize 32 A800 GPUs and approximately 200 million query-item samples.  
Task types include: Query Generation Task: Input item information, the model must generate its associated query. During training, one query is randomly selected from the multiple queries co-occurring with the item as the label.
Item Cloze Task: Randomly masks 1-3 cloze spans within the item description text, each span being 1-3 characters long. The model predicts the masked content.
Product Suffix Completion Task: Randomly splits the item description text into a prefix and a suffix. Input is a random product query and the item prefix, requiring the model to generate and complete the suffix.
Within each batch, the three tasks are randomly assigned in a 1:1:1 ratio and identified by special string tokens.

\textbf{GSID Training Phase}

Building upon the fully pre-trained mengzi\_t5 base model, training proceeds incrementally across 4 steps, with each step corresponding to a codebook size of 128. For any given step, samples within a batch are formed by concatenating several sample groups sharing identical prefix codes. Samples within a group represent hard (difficulty-increasing) samples, while samples across groups represent easy samples. We perform a gather operation across different GPUs to compute the contrastive loss.
Prior to the main training within each step, an initial phase of n batches is dedicated solely to contrastive loss training. During this phase, only the contrastive loss is used to update the model, aiming to achieve a reasonable spatial distribution of query-item vectors. Subsequently, we further optimize the Query-Item Alignment Objective and Code Commitment Objective.
The entire GSID training utilizes 32 A800 GPUs and approximately 200 million query-item samples. Each step is trained for only 1 epoch.

\end{document}